\documentclass[11pt]{article}
\usepackage{hyperref}
\pdfoutput=1
\usepackage{natbib} % this is needed for JFM bibliography style
\usepackage[]{graphicx}    % this is required to include eps files
\usepackage[]{color}       % enables color
\usepackage{ulem} % required to xout text
\include{psfig}
\newcommand{\fn}[1]{\footnote{#1}}
\newcommand{\lbr}{\linebreak} 
\newcommand{\bc}{\begin{center}} 
\newcommand{\ec}{\end{center}} 
\newcommand{\bslide}{\begin{slide}} 
\newcommand{\eslide}{\end{slide}}

\newcommand{\be}{\begin{equation}} 
\newcommand{\ee}{\end{equation}} 
\newcommand{\bml}{\begin{multline}} 
\newcommand{\eml}{\end{multline}}
\newcommand{\bspl}{\begin{split}} 
\newcommand{\espl}{\end{split}}
\newcommand{\bal}{\begin{align}} 
\newcommand{\eal}{\end{align}}
\newcommand{\bquote}{\begin{quote}}
\newcommand{\equote}{\end{quote}}
\newcommand{\bd}{\begin{displaymath}} 
\newcommand{\ed}{\end{displaymath}} 
\newcommand{\bea}{\begin{eqnarray}} 
\newcommand{\eea}{\end{eqnarray}} 
\newcommand{\bda}{\begin{eqnarray*}} 
\newcommand{\eda}{\end{eqnarray*}} 
\newcommand{\ba}{\begin{array}} 
\newcommand{\ea}{\end{array}} 
\newcommand{\bit}{\begin{itemize}} 
\newcommand{\eit}{\end{itemize}}
\newcommand{\ben}{\begin{enumerate}} 
\newcommand{\een}{\end{enumerate}}
\newcommand{\bt}{\begin{tabular}} 
\newcommand{\et}{\end{tabular}} 
\newcommand{\beab}{\begin{abstract}} 
\newcommand{\eab}{\end{abstract}} 
\newcommand{\ol}{\overline} 
\newcommand{\pa}{\partial}

\newcommand{\mean}[1]{\left\langle #1 \right\rangle}
\newcommand{\lam}{\lambda}
\newcommand{\del}{\delta}
\newcommand{\Del}{\Delta}
\newcommand{\Reth}{Re$_\theta$}
\newcommand{\utau}{u_\tau}

\usepackage{amssymb,amsmath,amsfonts,amsbsy}
\usepackage{graphicx}
\usepackage{subfigure}
\usepackage{epsfig}
\usepackage{tabularx}
\usepackage[english]{babel}
\usepackage{wrapfig}
\usepackage{color}
\usepackage{supertabular}
\begin{document}
\title{LES of an Inclined Jet into a \\ Supersonic Turbulent Crossflow}
\author{Antonino Ferrante$^1$, Georgios Matheou$^2$, Paul E. Dimotakis$^2$,\\
Mike Stephens$^3$, Paul Adams$^3$, Richard Walters$^3$
\\ \\
$^1$ Aeronautics \& Astronautics, \\ 
\vspace{6pt} University of Washington, Seattle, WA 98195, USA \\
$^2$ Graduate Aeronautical Laboratories, \\ 
\vspace{6pt} California Institute of Technology, Pasadena, CA 91125, USA \\
$^3$ Data Analysis and Assessment Center, U.S. Army \\
Engineer Research and Development Center, MS 39180, USA}
\maketitle
%% The abstract (in this file, and that submitted as text to arXiv) should include the exact phrase
%% "fluid dynamics video" or "fluid dynamics videos"
\vspace{-.8cm}
\begin{abstract}
This short article describes flow parameters, numerical method, and animations of the fluid dynamics video ``LES of an Inclined Jet into a Supersonic Turbulent Crossflow" (\href{http://ecommons.library.cornell.edu/bitstream/1813/14073/3/GFM-2009.mpg}{high-resolution} and 
\href{http://ecommons.library.cornell.edu/bitstream/1813/14073/2/GFM-2009-web.m1v}{low-resolution} video). 
We performed large-eddy simulation with the sub-grid scale (LES-SGS) stretched-vortex model of momentum and scalar transport to study the gas-dynamics interactions of a helium inclined round jet into a supersonic ($M=3.6$) turbulent (\Reth$\ =13\times10^3$) air flow over a flat surface. 
The video shows the temporal development of Mach-number and magnitude of density-gradient in the mid-span plane, and isosurface of helium mass-fraction and $\lam_2$ (vortical structures). 
The identified vortical structures are sheets, tilted tubes, and discontinuous rings. The vortical structures are shown to be well correlated in space and time with helium mass-fraction isosurface ($Y_{\rm He}=0.25$).
\end{abstract}
\vspace{-.6cm}
\section{Flow parameters}
\begin{figure*}[h]
\centering
\includegraphics[width=.9\textwidth]{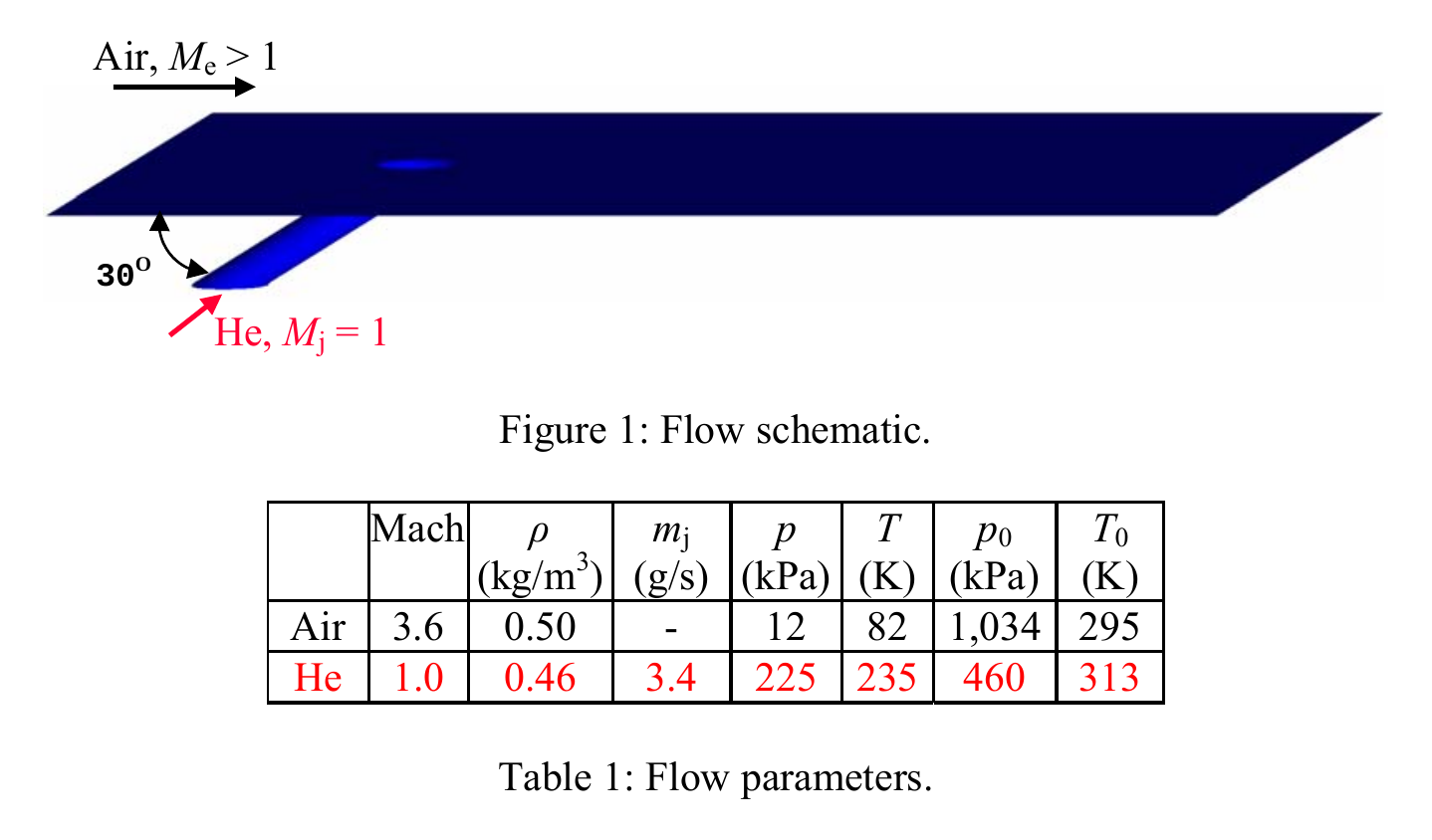} %%% NOTE: to make it work do not put extension .eps !!!
\label{fig:jetmodel}
\vspace{-1cm}
\end{figure*}

Figure \ref{fig:jetmodel} shows the flow schematic. Helium is injected through an inclined round jet into a supersonic turbulent air flow over a flat surface. 
In the present investigation, the jet axis forms a 30$^\circ$ angle with the streamwise direction of the air flow. 
The flow parameters of air and helium are reported in Table \ref{fig:jetmodel}. 
The jet diameter, $d$, is 3.23$\times 10^{-3}$~m, and the boundary layer thickness, $\del$, of the air flow is 2$\times10^{-2}$~m as in the experimental study of \cite*{mad06}. The air free-stream Mach number is 3.6, the jet Mach number is 1.0, and the jet to free-stream momentum ratio, $\ol q=(\rho U^2)_{\rm j}/(\rho U^2)_{\rm e}$, is 1.75. 
The Reynolds number of the air flow based on the momentum thickness is ${\sl Re}_\theta=U_{\rm e} \theta/\nu_{\rm w}=13 \times 10^3$ (${\sl Re}_\del=U_{\rm e} \del/\nu_{\rm w}=113\times10^3$), where $U_{\rm e}$ is the free-stream air velocity and $\nu_{\rm w}$ is the kinematic viscosity of air computed at the wall for adiabatic wall conditions.
%%%%%%%%%%%%%%%%%%%%%%%%%%%%%%%%%%%%%%%%
\vspace{-.2cm}
%%%%%%%%%%%%%%%%%%%%%%%%%%%%%%%%%%%%%%%%
\vspace{-.2cm}
\section{Numerical method}
Large-eddy simulation was performed using the sub-grid scale (LES-SGS) stretched-vortex model of momentum and scalar transport developed by Pullin and co-workers \cite[]{pullin97,pullin00,pullin00-1,pullin02}. 
A hybrid numerical approach  \cite[]{hill04,pantano07} with low numerical dissipation that uses tuned centered finite differences (TCD) in smooth flow regions, and a weighted essentially non-oscillatory (WENO) scheme \cite[]{osher94,shu96} around discontinuities and ghost-fluid boundaries is employed. 
The level-set approach with the ghost-fluid method \cite[]{fedkiw99} is used to treat the complex boundary (Fig.~\ref{fig:jetmodel}) where no-slip and adiabatic boundary conditions are applied. 
The framework for block-structured Adaptive Mesh Refinement in Object-oriented C++ (AMROC) by \cite{deiterd03} was adopted. 
The computational domain is a parallelepiped with sides $-0.025~\mbox{m}\le x \le 0.0774$~m, $-0.006~\mbox{m}\le y \le 0.0452$~m, and $-0.0256~\mbox{m}\le z \le 0.0256$~m (i.e., $-7.74\le x/d \le 24$, $-1.86\le y/d \le 14$, and $-7.92 \le z/d \le 7.92$), in the streamwise, wall-normal and spanwise directions, respectively. The flat-wall boundary is the $y=0$ plane and the center of the jet-exit plane is at the axes origin (Fig.~\ref{fig:jetmodel}). 
The basic mesh is $256 \times 128 \times 128=4.2\times10^6$ grid points with a base mesh spacing of $\Del x_{\rm b}=4\times10^{-4}$~m. The dynamic mesh refinement (AMR=2) adds one level of finer mesh on the coarse mesh, increasing the effective total number of mesh points to approximately $16\times10^6$ with a fine-mesh spacing of $\Del x_{\rm f}=2\times10^{-4}$~m. The use of AMR=2 saves about half grid points with respect to the fully refined mesh of $33\times10^6$ grid points. 
The Kolmogorov length-scale, $\eta$, of the shear layer created between the high-speed helium jet and low-speed air stream is of about 1.5~$\mu$m, i.e., $\eta$ is of $O(100)$ times smaller than the fine-mesh spacing. The SGS-TKE is less than 20\% of the total TKE in most of the flow field . Thus, the resolution criterion for a sufficiently resolved LES-SGS \cite[]{pope04} is satisfied.
\vspace{-.3cm}
\subsection{Inflow conditions}
The transition and spatial development of the helium jet were found to be strongly dependent on the inflow conditions of the crossflow \cite[]{af-etal-aiaa09}. These results indicate that correct turbulent inflow conditions are {\em necessary} to predict the main flow characteristics, dispersion and mixing of a gaseous jet in a supersonic turbulent crossflow. 
% Objective
A methodology for the generation of synthetic turbulent inflow conditions for LES of spatially developing, supersonic, turbulent wall-bounded flows has been developed by \cite{af-etal-aiaa10}. 
A brief description is here given. Inflow conditions are computed as the sum of zero-pressure gradient, compressible turbulent-boundary-layer (ZPG-CTBL) mean profiles and turbulence fluctuations. First, the friction velocity, $\utau$, and velocity profile \cite[]{vk30} %,col56} 
of the incompressible ZPG-TBL are computed by prescribing the Reynolds number based on boundary-layer thickness, $Re_\del$ \cite[]{af-se-jcp04}.
%, as described in Appendix A of Ferrante \& Elghobashi (2004) \cite{af-se-jcp04}. 
The resulting velocity profile is then transformed into the velocity profile of the ZPG-CTBL at $M_{\rm e}=3.6$ according to van Driest \cite[]{vd51,smits06,knight06}. 
Last, the temperature profile of the ZPG-CTBL is computed using the Walz formula \cite[]{walz69} assuming an adiabatic wall. 
Inflow turbulence fluctuations are generated by modifying the methodology of \cite[]{af-se-jcp04} to supersonic flows at high-Reynolds number. 
A model spectrum \cite[]{pope00}of turbulence kinetic energy, $E(k)$, and the Reynolds stresses, $\mean{u_i u_j}^+$, are prescribed at the inflow plane.
The Reynolds stresses at $M_{\rm e}=3.6$ are obtained by scaling the incompressible Reynolds stresses  of a ZPG-TBL at \Reth=2900 obtained via DNS \cite[]{af-se-jfm05} with the theoretical local profile $\rho/\rho_{\rm w}$. 
Such scaling is justified by the DNS results of \cite{guarini00}. 
\newpage
\section{Animations}
The fluid dynamics video ``LES of an Inclined Jet into a Supersonic Turbulent Crossflow" (\href{http://ecommons.library.cornell.edu/bitstream/1813/14073/3/GFM-2009.mpg}{high-resolution} and 
\href{http://ecommons.library.cornell.edu/bitstream/1813/14073/2/GFM-2009-web.m1v}{low-resolution} video) shows five animations (A1 to A5)\fn{Each animation lasts about 23~s and shows 30~frames/s for a total of 696 frames. The temporal development of the flow is shown for about $10^{-3}$~s. Thus, the flow is shown 23,000 times slower than at its actual speed. 
}:
\ben
\item Mach-number, $M$, contours in the mid-span plane;
\item contours of density-gradient magnitude, $|\nabla \rho|$, in the mid-span plane;
\item isosurface of helium mass-fraction, $Y_{\rm He}=0.25$;
\item isosurface of $\lam_2$ (vortical structures)\fn{The vortical structures are educed using the $\lam_2$-method \cite[]{hus95}, where $\lam_2$ is defined as the second largest eigenvalue of the tensor $(S_{ik}S_{kj}+\Omega_{ik}\Omega_{kj})$, where $S_{ij}\equiv(\pa_jU_i+\pa_iU_j)/2$ is the strain rate tensor, and $\Omega_{ij}\equiv(\pa_jU_i-\pa_iU_j)/2$ is the rotation rate tensor.};
\item overlapped isosurface of helium mass-fraction $Y_{\rm He}=0.25$ (yellow), and vortical structures (blue).
\een
The Mach-number contours (A1) show that boundary-layer turbulence/\lbr bow-shock interaction results in shock-wave unsteadiness that affects the roll-up of the shear-layer (formed between air stream and helium jet) and, consequently, modulates in time the size and shape of the barrel-shock (black region near the jet exit). 
%The shear-layer formed between air stream and helium jet is clearly visible as a black solid wiggly line in the contours of the density-gradient magnitude.
The contours of the density-gradient magnitude (A2) show the bow-shock and the shear-layer formed in between the air stream after the bow-shock and the expanded helium jet. Both contours show large-scale structures advected downstream. Jet unsteadiness, lateral and wall-normal helium dispersion, and three-dimensional structure of the helium jet are shown in the animation of $Y_{\rm He}=0.25$ isosurface (A3). 
The vortical structures (isosurface of $\lam_2$ shown in A4) are  {\em sheets} near the jet exit where the shear formed between the air-stream and the helium jet is large. Downstream the jet exit, the vortical structures are mostly {\em tilted tubes} which sometimes look like {\em discontinuous rings}.  
In A5, the isosurface of $Y_{\rm He}=0.25$ (yellow) mostly envelopes the isosurface of $\lam_2$ (blue). The two isosurfaces, showing helium-jet puffs and vortical structures, are well correlated in space and time. The vortical structures look like muscles that move the isosurface of helium mass-fraction, contributing to helium dispersion and helium mass-fraction convoluted isosurface.
\subsection*{Acknowledgments}
This work was supported by AFOSR Grants FA9550-04-1-0020 and FA9550-04-1-0389, and by Caltech funds. The authors would like to thank Dr. C. Pantano for support with the AMROC software, the Center of Advanced Computing Research (CACR) at Caltech for computing time.
%
%\bibliography{C:/AF-Files/_refs_newcomm_psfig/ref_total}
\bibliographystyle{jfm}

\end{document}